\documentclass[12pt,letterpaper]{article}

\usepackage{epsfig}
\usepackage{amsmath}
\usepackage{amssymb}
\usepackage{amsthm}
\usepackage{indentfirst}
\usepackage{xspace}
\usepackage{multirow}
\usepackage{hyperref}
\usepackage{xcolor}
\definecolor{darkred}{rgb}{0.5,0.15,0.15}
\hypersetup{colorlinks=true,urlcolor=darkred,linkcolor=darkred,citecolor=darkred}
\usepackage{verbatim}
\usepackage[letterpaper,margin=1in,headheight=15pt]{geometry}
\usepackage{mathpazo}
\usepackage{authblk}
\usepackage{tikz-cd}
\usepackage{capt-of}
\usepackage{silence}
\usepackage[missing=]{gitinfo2}
\usepackage{calc}
\usepackage{lipsum}
\usepackage{braket}

\usepackage{tikz}
\usetikzlibrary{calc}  
\usetikzlibrary{topaths}
\usetikzlibrary{decorations}
\usetikzlibrary{decorations.pathmorphing}

\numberwithin{equation}{section}

\newcommand{\cR}{\mathcal{R}}

\newcommand{\la}{\langle}
\newcommand{\ra}{\rangle}

\newcommand{\bQ}{\overline{Q}}

\newcommand{\bSigma}{\overline{\Sigma}}
\newcommand{\oS}{\overline{\Sigma}}

\newcommand{\cM}{\ensuremath{\mathcal M}}
\newcommand{\cO}{\ensuremath{\mathcal O}}

\newcommand{\cX}{\ensuremath{\mathcal X}}

\newcommand{\R}{\ensuremath{\mathbb R}}

\newcommand{\Z}{\ensuremath{\mathbb Z}}

\newcommand{\half}{\ensuremath{\frac{1}{2}}}

\newcommand{\N}{{\mathcal N}}
\newcommand{\cN}{{\mathcal N}}

\newcommand{\I}{{\mathrm i}}
\newcommand{\E}{{\mathrm e}}
\newcommand{\de}{\mathrm{d}}

\newcommand{\cc}{\mathrm{cc}}
\newcommand{\cons}{\mathrm{cons}}
\newcommand{\bulk}{\mathrm{b}}
\newcommand{\defect}{\mathrm{d}}

\newcommand{\abs}[1]{\lvert#1\rvert}

\newcommand{\IP}[1]{\langle#1\rangle}

\newcommand{\eps}{\epsilon}

\newcommand{\bea}{\begin{eqnarray}}
\newcommand{\eea}{\end{eqnarray}}

\newcommand{\ti}[1]{\textit{#1}}

\newcommand{\fourd}{{\mathrm{4d}}}
\newcommand{\twod}{{\mathrm{2d}}}
\newcommand{\twodfourd}{{\mathrm{2d-4d}}}

\newcommand{\bG}{\overline{G}}
\newcommand{\overlineL}{\overline{L}}
\newcommand{\bJ}{\overline{J}}

\newcommand{\SD}{\mathrm{S}}

\newcommand{\ovS}{\overline{S}}

\DeclareMathOperator{\End}{End}
\DeclareMathOperator{\Def}{Def}

\DeclareMathOperator{\Vol}{Vol}

\begin{document}

\hfill UTTG-18-20

\vspace{1.5cm}

\begin{center}
{\LARGE{Deformations of surface defect moduli spaces}}
\end{center}

\begin{center}
{\large{Andrew Neitzke\footnote{Department of Mathematics, Yale University} and Ali Shehper\footnote{Department of Physics, University of Texas at Austin}}}
\end{center}

\vspace{0.7cm}

{{{\tiny \color{gray} \tt \gitAuthorIsoDate}}
{{\tiny \color{gray} \tt \gitAbbrevHash}}}

{\abstract{Given a 4d $\cN=2$ supersymmetric theory with an $\cN=(2,2)$ supersymmetric surface defect,
a marginal perturbation of the bulk theory induces a complex structure deformation of the defect moduli space. We describe a concrete way of computing this
deformation using the bulk-defect OPE.
}}

\setcounter{page}{1}

\section{Introduction}

It is by now a well known principle
that to get a complete picture of a quantum field theory one 
should study not only local operators, 
but more generally defects of all dimensions.
These defects give rise to rich algebraic, topological and geometric structures.
A deformation of the quantum field theory then must induce deformations of 
all of these structures, fitting together in a self-consistent way, which 
may be rather intricate in its full generality.

This paper concerns a small part of that story. Given an $\cN=2$ supersymmetric quantum
field theory in four dimensions, and an $\cN=(2,2)$ supersymmetric surface defect,
we ask, \ti{how does a deformation of the theory induce a deformation of the complex
structure of the surface defect moduli space?}

In the rest of this introduction 
we describe this problem more carefully and formulate our
proposed answer.

\subsubsection*{Bulk and defect moduli spaces for 4d $\cN=2$ theories}

Suppose we are given a $d=4$, $\N=2$ 
supersymmetric theory.
The theory has a moduli space $\cM_{\fourd}$ of marginal supersymmetric couplings,
which is naturally complex.
Now we introduce 
a $\frac12$-BPS surface defect preserving $d=2$, $\N=(2,2)$ supersymmetry
\cite{Alday:2009fs,Gaiotto:2009fs}.
Holding the 4d coupling $\tau \in \cM_{\fourd}$ fixed,
such a surface defect has a moduli space $\cM_{\twod}$ of marginal chiral deformations, which is again a complex manifold.

Letting both couplings vary, 
we have a combined 2d-4d moduli space $\cM_{\twodfourd}$ which is a holomorphic
fiber bundle over $\cM_{\fourd}$. As we vary $\tau \in \cM_{\fourd}$, the complex structure of the fiber $\cM_{\twod}$ may
in general vary.
The infinitesimal version of this statement is that there is a linear map
\begin{equation} \label{eq:def-map}
	T \cM_{\fourd} \to \Def(\cM_{\twod})
\end{equation}
where $\Def(\cM_{\twod})$ means the space of linearized deformations of the complex
structure of $\cM_{\twod}$.

\subsubsection*{The case of class $S$}

On abstract grounds we know that the map \eqref{eq:def-map} exists, but one might
wonder whether this map could be zero. Indeed in some examples it will be zero (e.g.
it must be zero if we study a ``surface defect'' which is just a $d=2$, $\cN=(2,2)$ theory
uncoupled from the bulk theory!)
Still, in some examples the map \eqref{eq:def-map} 
is known to be nonzero, as we now explain.

Suppose we consider a theory of class $S$ obtained by compactifying the 6d $(2,0)$ theory on a surface $C$.
In this case 
one can construct a surface defect by starting with a surface defect of the $(2,0)$ theory
and placing it at a point $z \in C$; these defects were introduced and studied in 
\cite{Alday:2009fs,Gaiotto:2009fs}.
For this surface defect we have (up to discrete covers)
\begin{equation}
	\cM_{\twod} = C.
\end{equation}
On the other hand, one of the essential insights of \cite{Gaiotto:2009we} was that 
(up to discrete covers) $\cM_{\fourd}$ is the Teichm\"uller space of $C$. It follows that
\begin{equation}
	T\cM_{\fourd} = \Def(C).
\end{equation}
Thus in this case we actually have
\begin{equation}
	T \cM_{\fourd} = \Def(\cM_{\twod});
\end{equation}
in other words, the map \eqref{eq:def-map} is not only nonzero but 
an isomorphism.

\subsubsection*{Computing the deformation intrinsically}

The question we address in this paper is: \ti{how can the map \eqref{eq:def-map} be understood and calculated intrinsically
in the language of QFT}, without relying on class $S$ descriptions or other features of specific examples?

Here is the answer we propose.
Let $R_{\fourd}$ denote the space of chiral operators of dimension $2$ in the bulk $\N=2$ theory.
Any operator $\Phi \in R_{\fourd}$ has a descendant $Q^4 \Phi$
which can be used to deform the theory 
in a way which preserves $d=4$, $\N=2$
supersymmetry and conformal invariance; this gives an identification
\begin{equation}
R_{\fourd} = T \cM_{\fourd}.
\end{equation}
Likewise, 
let $R_{\twod}$ denote the space of chiral local operators of dimension $1$ living on the surface defect.
Any operator $\Sigma \in R_{\twod}$ can be 
used to deform the surface defect in a way which preserves $d=2$, $\N=(2,2)$ supersymmetry
and conformal invariance; this gives an identification 
\begin{equation}
R_{\twod} = T \cM_{\twod}.
\end{equation}
Now suppose
given $\Phi \in R_{\fourd}$ and $\bSigma \in \overline{R_{\twod}}$.
We let $\mu_\Phi(\bSigma)$ denote the most singular term in the bulk-defect OPE,
\begin{equation} \label{eq:mu-phi}
	\Phi(x) \bSigma(0) = \frac{4}{\I \pi} \frac{\mu_\Phi(\bSigma)}{\abs{x}^2} + \cdots
\end{equation}
Counting R-charges and dimensions shows that $\mu_\Phi(\bSigma) \in R_{\twod}$.
Thus we have obtained a linear map
\begin{equation}
\mu_\Phi: \overline{R_{\twod}} \to R_{\twod}.
\end{equation}
Since $R_{\twod} = T \cM_{\twod}$,
such a map can be interpreted as
\begin{equation}
	\mu_\Phi: \overline{T \cM_{\twod}} \to T \cM_{\twod},
\end{equation}
or equivalently $\mu_\Phi \in \Omega^{0,1}(\cM_{\twod}, T \cM_{\twod})$.
Moreover, $\mu_\Phi$ obeys the condition $\bar\partial \mu_\Phi = 0$.
As we review in \autoref{app:beltrami}, the $\bar\partial$-cohomology
class of such a $\mu_\Phi$ determines an element
\begin{equation}
	[\mu_\Phi] \in \Def(\cM_{\twod}).
\end{equation}
Our main claim is that $[\mu_\Phi]$ represents the deformation of $\cM_{\twod}$
which is induced by perturbing the bulk theory using the operator $Q^4 \Phi$.

Our derivation of this claim is given in \autoref{sec:def4d} below.
It builds on the study of 
operator mixing in deformations of CFTs \cite{Seiberg:1988pf, Kutasov:1988xb, Ranganathan:1992nb, Ranganathan:1993vj}. It was argued in these works that as a CFT is deformed, the OPE 
between the perturbing operator and other local operators determines 
a connection on the vector bundle of local operators. This
connection is responsible for the phenomenon of mixing between local operators
as we move on the conformal manifold of the CFT. 
In a similar way, we find that the OPE between local operators in the 
bulk QFT and local operators inserted on the defect contains information 
about operator mixing between marginal chiral descendants $Q^2 \Sigma$
and marginal anti-chiral descendants $\bQ^2 \bSigma$ as we move on the moduli
space $\cM_\fourd$. This operator mixing is a manifestation of the deformation
of complex structure of $\cM_\twod$.

In \autoref{sec:ex} we discuss one concrete example, where the bulk theory
is the pure $\N=2$ theory with gauge group $U(1)$, and the defect is a supersymmetric
``solenoid.'' In this case the moduli space $\cM_{\twod}$ is a $1$-dimensional complex
torus, 
whose complex modulus is the 4d gauge coupling $\tau$.
We verify in this case that the OPE between the bulk deformation 
operator $\Phi = \phi^2$ and the defect anti-chiral operator $\bSigma = \overline\phi$
gives the Beltrami differential associated to the complex deformation
of $\cM_{\twod}$ as expected.
(We remark that this example
can be thought of as a class $S$ theory associated to the Lie algebra $gl(1)$
where $C$ is a torus.)

\subsubsection*{Comments and future directions}

\begin{enumerate}

	\item In this paper we examine one very specific deformation problem, that of moduli spaces
	of $\half$-BPS surface defects in 4d $\cN=2$ theories. The basic mechanism we find, that these
	deformations are controlled by the bulk-defect OPE, seems likely to recur in other dimensions
	and other amounts of supersymmetry. It
	would be interesting to explore other examples.

	\item Although our analysis is intended to apply to a general $\cN=2$ theory and surface
	defect, the only example we consider in detail is that
	of a free theory and surface defect. It would be interesting to verify our analysis
	directly in an interacting theory. For example, we could consider the pure $SU(2)$ theory 
	with $N_f = 4$. This theory has a class $S$ realization where $C$ is a four-punctured sphere,
	and a corresponding canonical surface defect with $\cM_{\twod} = C$;
	if we consider the bulk-defect OPE in this theory we thus expect to obtain a 
	Beltrami differential representing the deformation of the four-punctured sphere which 
	changes the cross-ratio of the punctures.

	\item One of the technical tools in our analysis is a computation of part of the covariant
	derivative of the conserved supercurrent of the 4d $\cN=2$ theory with respect to the coupling:
	we find that (at least with the regularization scheme we use) 
	the supercurrent $\bJ$ mixes with a descendant of the chiral perturbation $\Phi$,
	schematically $\nabla_\eps \bJ = Q^3 \Phi + \cdots$ (see \autoref{sec:J-deformations} for the precise statement).
	This kind of mixing might occur more generally for deformations of supersymmetric
	theories, and if so it could be interesting to study more systematically.

	\item In this paper we focus on marginal deformations of the bulk
	theory, descending from chiral operators $\Phi$ of dimension $2$.
	One could similarly consider $\Phi$ of dimension $2+k$ with $k>0$,
	which would give rise to irrelevant deformations 
	of the bulk theory. The bulk-defect
	OPE then gives a map from defect anti-chiral 
	operators of dimension $1$ to defect chiral operators of 
	dimension $1+k$. When $C$ is 1-dimensional, such a map
	can be interpreted as a \ti{higher Beltrami differential}
	on $C$, and thus irrelevant
	perturbations of the bulk theory correspond to perturbations of a
	\ti{higher complex structure} on $C$ in
	the sense of \cite{1812.11199}.
	It would be very interesting to understand the meaning of this
	higher complex
	structure on $C$ in terms of the physics of the surface defect; one intriguing possibility is to interpret it as a deformation
	of a larger moduli space including both marginal and irrelevant deformations of the defect.

\end{enumerate}

\subsection*{Acknowledgements}

We thank Chris Beem, David Ben-Zvi, Jacques Distler, Aaron Fenyes, Greg Moore and Kyriakos Papadodimas for helpful conversations.
The work of AN was supported by NSF grants DMS-2005312 and DMS-1711692. 
AN also thanks the Mathematical Sciences
Research Institute for hospitality during the fall 2019 semester, supported by NSF grant DMS-1440140.
The work of AS was supported by NSF grant DMS-1711692.

\section{Properties of 2d-4d systems} \label{sec:prop}

This section is divided as follows. In \autoref{sec:algebra}, we review the superconformal algebra of 2d-4d systems. Explicit commutation relations are not given here but can be found in \autoref{app:4d} and \autoref{app:2d}. In \autoref{sec:marg} we discuss supersymmetry-preserving marginal operators of 4d and 2d systems. 

\subsection{Superconformal symmetry of 2d-4d systems} \label{sec:algebra}

We start by reviewing the $\mathcal{N}=2$ superconformal algebra in four dimensions. Its bosonic subalgebra is $\mathfrak{so}(5,1) \oplus \mathfrak{su}(2)_r \oplus \mathfrak{u}(1)_r$ and the fermionic generators transform in a doublet of $\mathfrak{su}(2)_r$ as well as a spinor representation of $\mathfrak{so}(5,1)$. The $\mathfrak{u}(1)_r$ charges of these generators are given in the paragraph below. Under the decomposition $\mathfrak{so}(4) \oplus \mathfrak{so}(1,1) \subset \mathfrak{so}(5,1)$, a Weyl representation of $\mathfrak{so}(5,1)$ decomposes into Weyl representations of $\mathfrak{so}(4)$ distinguished by the action of $\mathfrak{so}(1,1)$. The generators with eigenvalues $+\frac{1}{2}$ and $-\frac{1}{2}$ are called Poincar\'e and conformal supercharges respectively. 

Using the isomorphism $\mathfrak{so}(4) \cong \mathfrak{su}(2)_L \oplus \mathfrak{su}(2)_R$, we denote the Poincar\'e supercharges as $Q^i{}_\alpha$, $\bQ_{j\dot{\alpha}}$ and the conformal supercharges as  $S_i{}^\alpha$ and $\overline{S}^{i\dot{\alpha}}$. Here $\alpha, \dot{\alpha}$ and $i$ index components in the fundamental representations of $\mathfrak{su}(2)_L$, $\mathfrak{su}(2)_R$ and $ \mathfrak{su}(2)_r$ respectively. Under the action of $\mathfrak{u}(1)_R$, $Q$'s and $\overline{S}$'s carry charge $+1$ while $S$'s and $\bQ$'s carry charge $-1$. 

In the presence of a surface defect $\SD$ on a plane $P$, the algebra of conformal symmetries $\mathfrak{so}(5,1)$ is reduced to $\mathfrak{so}(3,1) \oplus \mathfrak{so}(2)_\perp$. Here $\mathfrak{so}(2)_\perp$ is the algebra of rotation symmetry in the plane orthogonal to $P$. If $P$ spans $\mathbb{R}^2_{x^1=x^2=0}$, then, with the conventions listed in Appendix \ref{app:4d}, the list of  fermionic symmetries is given in Table \ref{table:nonlin}.

The unbroken symmetries generate the two-dimensional $(2,2)$ superconformal algebra as is evident by studying the action of unbroken bosonic symmetries on the supercharges. The $\mathfrak{u}(1)_r$ symmetry of $\N=2$ algebra is preserved; its charge is identified with half the charge of $\mathfrak{u}(1)_V$ R-symmetry. The $\mathfrak{su}(2)_R$ symmetry is explicitly broken to its Cartan, generated in our notation by $R_3$. The sub-algebra of $\mathfrak{su}(2)_R \oplus \mathfrak{so}(2)_\perp$ generated by $4 R_3 + 2 J_{12}$ is naturally identified with the axial R-symmetry in two dimensions. Finally, the $\mathfrak{u}(1)$ algebra generated by $J_{12} + R_3$ is the commutant of the embedding.

\begin{table}[ht] 
\centering 
\begin{tabular}{| c | c | c | c | c | c | c | c | c | c | c |} 
\hline
4d notation & $Q^1{}_1$ & $Q^2{}_2$ & $\bQ_{1\dot{1}}$ & $\bQ_{2\dot{2}}$ &  $S_1{}^1$  &$S_2{}^2$ &  $\overline{S}^{1\dot{1}}$   & $\overline{S}^{2\dot{2}}$ \\ [0.5ex] 
\hline
2d notation & $G^+_{-1/2}$ & $\bG^-_{-1/2}$ & $G^-_{-1/2}$ & $\bG^+_{-1/2}$ &  $G^-_{1/2}$  &$\bG^+_{1/2}$ &  $G^+_{1/2}$   & $\bG^-_{+1/2}$ \\ [0.5ex] 
\hline 
$\mathfrak{u}(1)_V$ & $+1$ & $+1$ & $-1$ & $-1$ & $-1$ & $-1$ & $+1$  & $+1$  \\ 
\hline
$\mathfrak{u}(1)_A$ & $+1$ & $-1$ & $-1$ & $+1$ & $-1$ & $+1$ & $+1$  & $-1$  \\
\hline
$\mathfrak{so}(2)_{34}$ & $+\frac{1}{2}$ & $-\frac{1}{2}$ & $+\frac{1}{2}$ & $-\frac{1}{2}$ &$-\frac{1}{2}$ & $+\frac{1}{2}$ &  $-\frac{1}{2}$ & $+\frac{1}{2}$  \\ 
\hline 
\end{tabular}
\caption{The supercharges preserved in the presence of a surface defect and their charges under $\mathfrak{u}(1)_V$, $\mathfrak{u}(1)_A $ and $\mathfrak{so}(2)_{34}$. The normalization of 2d operators in terms of the 4d notation contains factors of $\frac{1}{2}$ (see equations (\ref{eq:B.9}), (\ref{eq:B.10}).) }
\label{table:nonlin} 
\end{table}

\subsection{Descendants of superconformal primaries} \label{sec:marg}

In this subsection, we discuss certain supersymmetry-preserving marginal  operators of 4d $\N = 2$ and 2d $\N=(2,2)$ systems. The operators that we discuss are distinguished by the properties that they are supersymmetric descendants of primary operators, and that they are marginal. 
\newline
\newline
\textit{Descendants of primaries in 4d $\N =2$ systems}
\newline

\indent A \textit{superconformal primary} is a local operator annihilated by the conformal supercharges $S_i{}^\alpha$ and $\overline{S}^{i\dot{\alpha}}$. If this operator is further annihilated by the right-handed (left-handed) Poincar\'e supercharges it is called a \textit{chiral (anti-chiral) primary}. If we assume that such an operator is a Lorentz scalar, its scaling dimension and $\mathfrak{u}(1)_r$ charge satisfy the equality 
\bea
\Delta = \mp r
\eea
where the upper (lower) sign holds for chiral (anti-chiral) primaries.

We obtain supersymmetry-preserving marginal operators through the action of left-handed (right-handed) Poincar\'e supercharges on chiral (anti-chiral) primaries of scaling dimension $\Delta = 2$. We denote these primaries as $\Phi_k$ ($\overline{\Phi}_k$); and the marginal operators as $Q^4 \Phi_k$ ($\bQ^4 \overline{\Phi}_k$).
\bea \label{eq:mo}
Q^4 \Phi_k &:=& \frac{\I}{64\pi} Q^1{}_1 Q^1{}_2 Q^2{}_1 Q^2{}_2 \Phi_k \\ \label{eq:mo2}
\bQ^4 \overline{\Phi}_k &:=& - \frac{\I}{64\pi} \bQ_{1\dot{1}} \bQ_{1\dot{2}} \bQ_{2\dot{1}} \bQ_{2\dot{2}} \overline{\Phi}_k
\eea 
Here $k$ goes from $1$ to $N$ --- the number of such operators in a theory. $Q^4 \Phi_k$ and $\bQ^4 \overline{\Phi}_k$ are complex conjugates in a spacetime with Minkowski signature; however, they are not complex conjugate in the Euclidean signature. This is because the Weyl spinors of $\mathfrak{so}(3,1)$ are complex conjugate while those of $\mathfrak{so}(4)$ are pseudo-real. In this paper, we will work in $\mathbb{R}^4$. 
\newline
\newline

\noindent\textit{Descendants of primaries in 2d $\N =(2,2)$ systems}
\newline

\indent In the notation of 2d $\N=(2,2)$ algebra the supercharges are denoted by $G^a_{r}$ and $\overline{G}^{a}_r$ where $a\in \{\pm\}$ and $r \in \{\pm \frac{1}{2}\}$ (see \autoref{app:2d}). The generators with $r=-\frac{1}{2}$ are the Poincar\'e supercharges while those with $r=+\frac{1}{2}$ are the conformal supercharges. Similar to the case of 4d $\N =2$, a \textit{superconformal primary} is defined as an operator that is annihilated by all of the conformal supercharges. As is well known (and is reviewed in Appendix \ref{app:2d}), the algebra of 2d $\N=(2,2)$ supersymmetry splits into a holomorphic and an anti-holomorphic sector. Hence the discussion of chiral and anti-chiral primaries also splits. 

In the holomorphic sector, the operators annihilated by $G^+_{-1/2}$ are called \textit{chiral primaries}, while the operators annihilated by $G^-_{-1/2}$ are called \textit{anti-chiral primaries}. The $L_0$ and $J_0$ eigenvalues of these operators satisfy 
\bea
l_0 = \pm \frac{j_0}{2} 
\eea
where the upper (lower) sign holds for chiral (anti-chiral) primaries. After taking into account the anti-holomorphic sector, the full $\N=(2,2)$ algebra admits four kinds of primary operators labelled as $(a,a)$, $(a,c)$, $(c,a)$ and $(c,c)$. We will also refer to these operators as twisted chiral, chiral, anti-chiral and anti-twisted chiral operators respectively.  

The supersymmetry preserving marginal operators are descendants of the primary operators of $(l_0, \overline{l}_0)$ weights equal to $(\frac{1}{2}, \frac{1}{2}) $, obtained by the action of those supercharges that do not annihilate them. For example, the descendant of a chiral primary $\Sigma (x)$ is
\bea 
Q^2 \Sigma = \frac{1}{2\sqrt{2}} Q^1{}_1 Q^2{}_2 \Sigma = \sqrt{2} G^+_{-1/2} \overline{G}^-_{-1/2} \Sigma
\eea
Similarly, the descendant of an anti-chiral primary $\oS$ is 
\bea \label{eq:defofoS}
\bQ^2 \oS = - \frac{1}{2\sqrt{2}} \bQ_{1\dot{1}} \bQ_{2\dot{2}} \oS = - \sqrt{2} G^-_{-1/2} \bG^+_{-1/2} \oS
\eea

\section{Deformations of 4d and 2d-4d systems} \label{sec:def4d}

\subsection{Deformation of pure 4d systems} \label{sec:pure-4d-def}

Consider deforming a 4d $\cN=2$ theory by a marginal chiral descendant,
\begin{equation}
	\delta S = \eps \int_{\R^4} \Phi^{(4)},
\end{equation}
where we defined 
\begin{equation}
\Phi^{(4)}(x) := Q^4 \Phi (x)\ \de^4 x; 
\end{equation}
the superscript in $\Phi^{(4)}$ denotes that this object is a 4-form. 
In this section we briefly review the effect of this deformation on the correlation functions. 
We use the notation that $\la \cdots \ra_{\epsilon}$ is the deformed correlation function while $\la \cdots \ra_{0}$ is the undeformed correlation function.

Formal path-integral manipulations would say that
the first derivative of $\la \cdots \ra_{\epsilon}$ with respect to 
$\epsilon$ is
\bea \label{eq:deform-def}
\frac{\de}{\de\epsilon } \la \prod\limits_{i=1}^n \cO_i (x_i) \ra_{\epsilon} \bigg|_{\epsilon=0} &=& - \int\limits_{\mathbb{R}^4}  \la \Phi^{(4)}(x) \prod\limits_{i=1}^n \cO_i (x_i) \ra_0
\eea
As it is written, the right side of \eqref{eq:deform-def} is potentially ill-defined, as there may be a short-distance singularity as $x \to x_i$. In order to cure this problem, one can follow a 4d version of a regularization procedure discussed e.g. in \cite{Ranganathan:1993vj} in the 2d context.\footnote{In particular, the regularization procedure discussed here corresponds to the $\overline{c}$-connection in \cite{Ranganathan:1993vj}.} That procedure entails cutting out small balls around the points of insertion of local operators, computing the integrated correlation function as a function of the radii of these balls, and dropping 
divergent terms as the radii are taken to zero.\footnote{An alternative approach to regularization is through the addition of contact terms in the OPE of the operators whose points of insertion coincide. This method is discussed in \cite{Seiberg:1988pf, Kutasov:1988xb}, and as explained in footnote 10 of \cite{Papadodimas:2009eu}, is equivalent to the procedure of  \cite{Ranganathan:1993vj}.} 

With this sort of regularization understood, \eqref{eq:deform-def} computes
the correlation functions of 
a family of 4d $\cN=2$ theories, to first order around a point of $\cM_{\fourd}$.
By construction the theories at $\eps = 0$ and $\eps \neq 0$ have the same 
vector space of local operators, since in \eqref{eq:deform-def}
the insertions in the correlation functions $\IP{\cdots}_\eps$ are drawn
from the original space of local operators, even when $\eps \neq 0$. 

So far, so good, to first order in $\eps$.
When one tries to go to higher order, however, the situation becomes 
more complicated. Following \cite{Papadodimas:2009eu}, one can take some inspiration from the case of 2-dimensional
field theories, discussed at some length in \cite{Ranganathan:1993vj}
following previous works including \cite{Seiberg:1988pf, Kutasov:1988xb, Ranganathan:1992nb}  (see also
\cite{Moore:1993zc}). The results of \cite{Papadodimas:2009eu} 
have been further developed and applied to various aspects
of 4d $\cN=2$ theories,
e.g. \cite{Baggio:2014ioa,Gerchkovitz:2016gxx,Baggio:2017aww,Niarchos:2018mvl}.
The expected picture 
can be summarized as follows: the space $\cM_{\fourd}$ carries a vector bundle whose fiber 
over a point $\tau \in \cM_{\fourd}$ is the space of local operators of the theory
with coupling $\tau$, but this vector bundle is not naturally trivial.
The fact noted above, that we can identify operators at $\eps = 0$ and $\eps \neq 0$ 
to first order in $\eps$, 
means that the bundle of local operators carries a connection 
$\nabla$. The connection $\nabla$ depends on the 
regularization scheme, and 
may have curvature in general, which is one manifestation of the 
phenomenon of operator mixing under marginal perturbations.

Revisiting \eqref{eq:deform-def}
from this perspective, we see that the appropriate interpretation is that this 
equation is true when the $\cO_i$ are sections
of the bundle of local operators obeying $\nabla_\eps \cO_i = 0$; 
this is the invariant way of 
saying the operator insertions are ``independent of $\eps$.''
More generally, if the $\eps$ dependence of the operators 
$\cO_i$ is chosen arbitrarily, \eqref{eq:deform-def} is replaced by
\begin{equation} \label{eq:deform-def-general}
\frac{\de}{\de\epsilon } \la \prod\limits_{i=1}^n \cO_i (x_i) \ra_{\epsilon} \bigg|_{\epsilon=0} = -  \int\limits_{\mathbb{R}^4}  \la \Phi^{(4)}(x) \prod\limits_{i=1}^n \cO_i (x_i) \ra_0 \, + \, \sum_{j=1}^n \la \nabla_\eps \cO_j(x_j) \prod_{i \neq j, i = 1}^n \cO_i(x_i) \ra_0 \, .
\end{equation}

\subsection{Deformations of 2d-4d systems}

Now we consider adding a $\half$-BPS surface defect to the system and deforming
again by a bulk 4d chiral descendant. 
We assume that the surface defect continues to be
$\half$-BPS in the deformed theory.

Working formally, as above,
deformed correlation functions to first order
can be computed by the analogue of \eqref{eq:deform-def},
\begin{equation}
\label{eq:deform-def-2d4d}
\frac{\de}{\de \epsilon} \la \SD(P) \prod\limits_{i=1}^n \cO^\bulk_i (x_i) \prod\limits_{i=1}^k \cO^\defect_i (y_i) \ra_{\epsilon} \bigg|_{\epsilon=0} = - \int\limits_{\mathbb{R}^4}  \la \Phi^{(4)}(x) \SD(P) \prod\limits_{i=1}^n \cO^\bulk_i (x_i) \prod\limits_{i=1}^k \cO^\defect_i (y_i) \ra_0 \, ,
\end{equation}
where $\SD(P)$ denotes the surface defect inserted on the plane $P$,
bulk local operators are inserted
at points $x_i \in \R^4$, defect local operators at points $y_i \in P$. 
The equation \eqref{eq:deform-def-2d4d} 
needs to be understood as including a regularization, e.g.
cutting out a tubular neighborhood of the defect and taking its
radius to zero with divergent terms dropped, in parallel to what we discussed for the bulk local operators.\footnote{
One might
have imagined a more general situation, namely that
the deformed correlation functions given 
by \eqref{eq:deform-def-2d4d} are
not those of an $\cN=(2,2)$ supersymmetric theory, but they
become supersymmetric after
shifting the action on the surface defect
by a marginal operator. However, all of
the marginal operators in an $\cN=(2,2)$
superconformal theory actually correspond to supersymmetry-preserving
deformations. Thus, if the theory is supersymmetric
with such an addition, it is also supersymmetric without one;
this means we can restrict our attention to the case where we do
not make such an addition.
}

To go beyond first-order deformations, 
as before, one needs to be more careful.
While we have not developed the theory in a fully systematic way, we
propose the following picture, parallel to the picture for 4d
local operators which we reviewed above.

First, given a regularization scheme
as we sketched above, we should obtain an
Ehresmann connection in the fiber bundle $\cM_{\twodfourd} \to \cM_{\fourd}$; this Ehresmann
connection defines what it means for the surface defect to be
``independent of $\eps$.''
As with the connection on local operators, this Ehresmann 
connection in $\cM_{\twodfourd}$ may have curvature, in the sense
that homotopic paths on $\cM_{\fourd}$ may lift to different
paths on $\cM_{\twodfourd}$. We will not 
need to consider that curvature here.

Second, there is a vector bundle over $\cM_{\twodfourd}$ consisting of
defect local operators. This vector bundle should carry a connection,
which defines what it means for defect local operators
to be ``independent of $\eps$.'' To describe this connection concretely 
we may use the Ehresmann connection
to make a splitting between the horizontal and vertical directions
in the fiber bundle $\cM_{\twodfourd}$.
Along the horizontal lifts of 
tangent vectors from the base $\cM_{\fourd}$ to $\cM_{\twodfourd}$, 
the connection 
is determined by the regularized integrals \eqref{eq:deform-def-2d4d},
which give us the first-order identification between spaces of 
local operators on the defect.
In this paper we will not need to discuss the connection in 
the vertical directions, but we remark that it is determined by
the same procedure we used in \autoref{sec:pure-4d-def},
now applied to a marginal perturbation of a 2d theory 
instead of a 4d theory.

We assume that the regularization scheme can be set
up in such a way that 2d and 4d perturbations commute with one
another. As a practical matter this means the following.
As we have already discussed, there is a connection
in the bundle of local operators over $\cM_{\twodfourd}$.
Because the vertical tangent bundle to $\cM_{\twodfourd}$
is identified with the space of marginal operators on the defect,
this gives a connection in the vertical tangent bundle.
On the other hand, the Ehresmann connection allows us to identify
nearby fibers and thus also induces
a connection in the vertical tangent bundle.
Our assumption is that these two connections are equal.

In parallel to the pure 4d case above, we interpret 
\eqref{eq:deform-def-2d4d} as giving the variation of
the correlation functions when all of the operators $\SD(P)$, 
$\cO^\bulk_i(x_i)$, $\cO^\defect_i(y_i)$ are deformed in a covariantly
constant fashion. In other words, 
the bulk local operators $\cO^\bulk_i$ are 
covariantly constant for the connection on bulk local operators over $\cM_{\fourd}$,
$\SD(P)$ is deformed along
the horizontal lift of a path from $\cM_{\fourd}$ to $\cM_{\twodfourd}$,
and the defect local operators $\cO^\defect_i$ are covariantly
constant for the connection on defect local operators over $\cM_{\twodfourd}$, evaluated
on the lifted path.

\subsection{Deformations of the supercurrent} \label{sec:J-deformations}

In a 4d $\cN=2$ theory the supersymmetry transformations $Q^I_\alpha$ and
$\bQ_{I\dot\alpha}$ are generated by corresponding conserved supercurrents
$J^I_\alpha$ and $\bJ_{I\dot\alpha}$ (we suppress the vector index on the currents).
We argue in \autoref{sec:deriv-J} below 
that the variation of the supercurrent is given by\footnote{Here and below, the covariant derivative $\nabla_\epsilon$ is understood to be computed at the point of the undeformed theory, i.e. $\epsilon=0$.}
\begin{equation} \label{eq:supercurrent-variation}
	\nabla_\eps \bJ_{I \dot\alpha} = (\Phi^{(3)})^\mu_{I \dot\alpha} + c \bJ_{I \dot\alpha}
\end{equation}
where $\Phi \in R_{\fourd}$ denotes the chiral operator corresponding
to the coupling $\eps$, 
$(\Phi^{(3)})^\mu_{I \dot\alpha}$ is the unique operator built by acting with
three $Q$'s on $\Phi$ which obeys
\begin{equation} \label{eq:def-phi3}
	\partial_\mu (\Phi^{(3)})^\mu_{I \dot\alpha} = \bQ_{I \dot\alpha} Q^4 \Phi,
\end{equation}
and $c$ is an undetermined constant.\footnote{This constant cannot be determined even in principle; the reason is that
by $\bJ$ we mean a section of the bundle of conserved supercurrents, and such 
a section is not quite unique; rather, it is determined up to $\bJ \to \E^{f(\eps)} \bJ$,
which would change $c \to c + \partial_\eps f$.}

In \cite{Papadodimas:2009eu} it is pointed out that the supercurrents form a 
bundle over $\cM_{\fourd}$ with a natural non-flat connection, which plays
an important role in a four-dimensional version of $tt^*$ geometry. 
That connection should be identified
with the \ti{projection} of $\nabla$ onto the bundle of supercurrents,
which has the effect of throwing away the $\Phi^{(3)}$ term above.
In contrast, for our purposes in the rest of this paper, the
$\Phi^{(3)}$ term will be the crucial part.

Below we will need to apply \eqref{eq:supercurrent-variation} in the situation
where a $\half$-BPS surface defect is inserted and $\bJ_{I \dot\alpha}$
is one of the supercurrents which is conserved in the presence of the defect. 
In this case one can
worry that \eqref{eq:supercurrent-variation} might need to be corrected 
to include a delta-function variation supported along the defect.
In \autoref{sec:deriv-J-with-defect} below we argue that this is not the case: \eqref{eq:supercurrent-variation} continues to hold even in the presence of the 
surface defect.

\subsection{Mixing between chiral and anti-chiral descendants}

We are now ready for our main computation. 
We consider the connection $\nabla$ acting on defect local operators.
This connection restricts to a connection 
on the bundle of marginal operators, and thus induces a connection
on vertical tangent vectors to $\cM_{\twodfourd}$. 
In turn, the bundle of vertical tangent vectors is decomposed into
holomorphic and antiholomorphic subbundles, corresponding to
chiral descendants $Q^2 \Sigma$ and anti-chiral
descendants $\bQ^2 \bSigma$ respectively.
What we will compute now is the mixing between the holomorphic and 
antiholomorphic subbundles as we move along $\cM_{\fourd}$.

To be precise,
let $\bSigma$ denote an
anti-twisted-chiral primary operator on
the surface defect.
We consider a first-order deformation of the bulk theory by a chiral
primary $\Phi$, lift it to a first-order path in $\cM_{\twodfourd}$ 
as we have been discussing above, and
extend $\bSigma$ to a section of the bundle of anti-twisted-chiral primaries
over this path. We also consider sections $J$, $\bJ$ of the bundle
of conserved supercurrents over the path, and use them to define the action of the
supercharges. 
Then
the main statement of this section is that, modulo operators 
which are $\bQ$-exact or $\bQ'$-exact, we have
\begin{equation} \label{eq:main-derivative}
	\nabla_\eps (\bQ^2 \bSigma) = Q^2 ( \mu_\Phi(\bSigma) ),
\end{equation} 
with $\mu_\Phi$ denoting the bulk-defect OPE we defined in \eqref{eq:mu-phi}.

The rest of this section is taken up with the derivation of \eqref{eq:main-derivative}, as follows.
We simplify notation a little by setting
	$\bQ := \bQ_{2 \dot{2}}$, $\bQ' := \frac{1}{2 \sqrt 2} \bQ_{1 \dot{1}}$,
so that $\bQ^2 = \bQ \bQ'$.
Now observe that modulo $\bQ$-exact terms we have
\begin{equation} \label{eq:interm}
	\nabla_\eps(\bQ^2 \bSigma) = \nabla_\eps (\bQ \bQ' \bSigma) = \nabla_\eps(\bQ) \bQ' \bSigma
\end{equation}
and
$\nabla_\eps(\bQ)$ is determined by the variation of the supercurrent,
given in
\eqref{eq:supercurrent-variation}. Continuing to work modulo $\bQ$-exact terms, this allows us to rewrite the RHS of \eqref{eq:interm} as
\begin{equation}
	\oint_\Gamma \Phi^{(3)}_{2 \dot{2}}(x) \bQ' \bSigma(0),
\end{equation}
where $\Gamma$ denotes any 3-cycle surrounding the point $x = 0$ where
the operator $\bSigma$ was inserted, and we have used the volume element of $\R^4$
to convert $\Phi^{(3)}_{2 \dot{2}}$ from a vector field to a 3-form 
(we will switch back and forth without further comment).
Modulo $\bQ'$-exact terms this is equal to
\begin{equation}
	\oint_\Gamma \bQ' \Phi^{(3)}_{2 \dot{2}}(x) \bSigma(0).
\end{equation}
All that remains is to evaluate this integral, for which
we can choose any convenient 3-cycle $\Gamma$. 
Let $(\rho,\psi)$ denote polar coordinates in the plane $P$ where the surface
defect lies, and $(r,\theta)$ denote polar coordinates in the orthogonal plane. 
We choose $\Gamma$ to be the boundary of the locus $\abs{r} \le R$,
$\abs{\rho} \le D$ for some $R, D > 0$. Thus $\Gamma = \cR \cup \cR'$ where
\begin{equation}
\cR = \{ \abs{r} = R, \abs{\rho} \le D \}, \qquad \cR' = \{ \abs{r} \le R, \abs{\rho} = D \}.
\end{equation}
In the limit $R \to 0$, the contribution from $\cR'$ vanishes.\footnote{
Note that this integral is convergent despite the presence of the surface
defect at $r = 0$; indeed the worst possible singularity
in the OPE between $\Phi^{(3)}_{2 \dot 2}$ and the surface defect
is of order $1/r$, as discussed in \autoref{sec:deriv-J-with-defect}, 
and this singularity is integrable.}
Thus we can get the correct answer by computing the integral only over $\cR$
for any fixed $D$ and then taking $R \to 0$.

Since the only bracket $\{ \bQ_{2\dot{2}}, \cdot \}$ containing $P_r$ 
is $\{\bQ_{2\dot{2}}, Q^2{}_1\} = 2 \E^{-\I \theta} \left(P_r - \frac{\I}{r} P_\theta\right)$, using \eqref{eq:mo} and \eqref{eq:def-phi3} gives
\begin{equation} \label{eq:phi3-r}
\Phi^{(3)}_r = -\frac{1}{32\pi} \E^{-\I\theta} Q^1{}_1 Q^1{}_2 Q^2{}_2 \Phi.
\end{equation}
Furthermore, the operator $\bQ' \Phi^{(3)}_r$ is a total derivative $\partial_\mu \cX^\mu$, because the anticommutator of $\bQ'=\frac{\bQ_{1\dot{1}}}{2\sqrt{2}}$ with $Q^1{}_1$ and $Q^1{}_2$ gives translations. Thus we have
\begin{equation}
 \oint_\cR \de^3 x \  \bQ' \Phi^{(3)}_r (x) \oS(0) =  \oint_\cR \de^3 x \  \partial_\mu \cX^\mu (x) \oS(0). \label{eq:def-int}
\end{equation}
We use the decomposition 
$\partial_\mu \cX^\mu = \partial_\theta \cX^\theta + \partial_\psi \cX^\psi + \partial_r \cX^r + \partial_\rho \cX^\rho$,
and consider the four terms in turn:
\begin{itemize}

\item The terms involving $\partial_\psi \cX^\psi$ and
$\partial_\theta \cX^\theta$ vanish since they are integrals of total derivatives over the circle.

\item The term with $\partial_\rho \cX^\rho$ reduces to the boundary 
integral $\oint_{S^1_R \times S^1_D} \cX^\rho$.
In the limit $R \to 0$, this boundary integral can be nonzero only if there is a singularity of order
$\frac{1}{r}$ in the OPE between $\chi^\rho$ and the surface defect; in the next
paragraph we show there is no such singularity.

Since the only bracket $\{\bQ_{1 \dot{1}}, \cdot \}$ containing
$P_\rho$ is $\{\bQ_{1 \dot{1}}, Q^1{}_1 \} = -2 \E^{-\I \psi} (P_\rho + \frac{\I}{r} P_\psi)$
we have
\begin{equation}
	\chi^\rho = \frac{1}{32\sqrt{2} \pi} \E^{-\I \psi-\I \theta} Q^1{}_2 Q^2{}_2 \Phi.
\end{equation}
Now we consider the OPE of $Q^1{}_2 Q^2{}_2 \Phi$ with the surface defect.
As is discussed in \autoref{sec:deriv-J-with-defect}, the most singular term in this
expansion has the form $\frac{\E^{\I \theta}}{r} Q^2{}_2 \xi$,
where $\xi$ is a superconformal primary operator on the defect obeying $Q^1{}_1 Q^2{}_2 \xi = 0$;
it follows that $Q^2{}_2 \xi = 0$, so the $\frac{1}{r}$ term vanishes, as desired.

\item Thus the only term that remains from \eqref{eq:def-int} is
\begin{equation}
	\oint_\cR \de^3 x \, \partial_r \cX^r(x) \oS(0).
\end{equation}
Since the only bracket $\{\bQ_{1\dot{1}}, \cdot \}$ containing $P_r$ is $\{\bQ_{1\dot{1}}, Q^1{}_2\} = 2 \E^{\I \theta} \left(P_r + \frac{\I}{r} P_\theta \right)$, we have
\begin{equation}
\cX^r = - \frac{\I}{32\sqrt{2} \pi} Q^1{}_1 Q^2{}_2 \Phi,
\end{equation}
so this integral becomes
\begin{equation} \label{eq:last-int}
\oint_\cR \de ^3 x \  \partial_r \cX^r (x) \oS(0)= - \frac{\I}{32\sqrt{2} \pi}\  Q^1{}_1 Q^2{}_2 \left( \oint_{\cR} \de^3 x \ \partial_r \Phi (x) \oS(0)  \right).
\end{equation}

\end{itemize}

To go further we replace $\Phi(x) \bSigma(0)$ by an OPE expansion
involving operators inserted at $0$.
The $(\Delta, r)$ quantum numbers of $\Phi $ and $\oS$ are $(2, -2)$ and $(1,1)$ respectively. Therefore, the most singular term in the OPE has $(\Delta, r) = (1,-1)$ and is hence a chiral operator on the defect, which 
we denote as $\mu_\Phi(\bSigma)$:
\bea \label{eq:OPE2}
	\Phi(x) \bSigma(0) = \frac{4}{\I \pi} \frac{\mu_\Phi(\bSigma)}{\abs{x}^2} + \cdots
\eea
Note that the only singularity in this OPE occurs at $x = 0$, the point of 
the surface defect where $\bSigma$ is inserted.
One might have worried that there would also be a singularity when $y$ lies at a general
point of the surface defect.
However, such a singularity does not occur, for the following reason.
Since $\Phi$ is a bulk chiral operator, its scaling dimension and $\mathfrak{u}(1)_r$ charge satisfy $\Delta = -r$. In a unitary theory, the bulk-to-defect OPE of $\Phi$ contains no singular terms, as the operators in such terms would have lower scaling dimension but the same $\mathfrak{u}(1)_r$ charge,
violating the unitarity bound $\Delta \ge \abs{r}$.

Now we are in position to calculate the integral:
\eqref{eq:last-int} becomes 
\begin{equation} \label{eq:eval}
	- \frac{\I}{16 \pi} \frac{4}{\I \pi} \left( \oint_\cR \partial_r \frac{1}{\abs{x}^2} \right) Q^2 \mu_\Phi(\bSigma)
\end{equation}
and we have
\begin{equation}
	\oint_\cR \de^3 x \, \partial_r \frac{1}{r^2 + \rho^2} = - 4 \pi^2 R \int_0^D \rho \de \rho \frac{2 R}{(R^2 + \rho^2)^2} = - 4 \pi^2 \frac{D^2}{R^2 + D^2}
\end{equation}
which in the limit $R \to 0$ becomes simply $- 4 \pi^2$, independent of $D$ as expected;
substituting this in \eqref{eq:eval} gives the final result
\begin{equation}
	 Q^2 \mu_\Phi(\Sigma)
\end{equation}
matching \eqref{eq:main-derivative} as desired. 

\subsection{The complex structure deformation}

Our main result \eqref{eq:main-derivative} expresses the phenomenon of mixing between
holomorphic and antiholomorphic tangent vectors to $\cM_{\twod}$
as we move along the lift of a path in $\cM_{\fourd}$. What remains is to explain
how this mixing is related to the infinitesimal deformation of complex structure
of $\cM_{\twod}$.

We first consider the properties of the tensor $\mu_\Phi \in \Omega^{0,1}(\cM_{\twod}, T \cM_{\twod})$
defined by \eqref{eq:OPE2}. 
We have (up to irrelevant constant factors)
\begin{equation}
	\IP{\Psi \vert \mu_\Phi(\bSigma)} = \abs{x}^2 \IP{\Psi \vert \Phi(x) \bSigma(0) } .
\end{equation}
In components on $\cM_{\twod}$, we could write this as
\begin{equation}
(\mu_\Phi)^b_{\bar a} g_{b \bar{b}} = \abs{x}^2 \IP{\bSigma_{\bar b} \vert \Phi(x) \bSigma_{\bar a}(0) } .
\end{equation}
By placing $x$ on the defect (using the fact that there is no singularity in
the bulk-defect OPE as noted above),
we can regard $\Phi$ as a chiral operator of the defect theory;
then making a conformal transformation we have
\begin{equation}
(\mu_\Phi)^b_{\bar a} g_{b \bar{b}} = 	\IP{\Phi(\infty) \bSigma_{\bar a}(1) \bSigma_{\bar b}(0)}.
\end{equation}
This is an \ti{anti-extremal correlator} on the defect,
since it involves one chiral operator and several anti-chiral
operators. It is a familiar fact from $tt^*$ geometry \cite{Cecotti:1991me} that these
correlators are covariantly antiholomorphic, 
and that the three-point correlator obeys 
an integrability condition, which in this case reads\footnote{In terms of the chiral ring coefficients $C_{IJ}^K$ in an $\cN=(2,2)$ theory, the integrability condition is usually written $\nabla_I C_{JK}^L = \nabla_J C_{IK}^L$. Using the metric to raise the $K$ and lower the $L$ indices it becomes $\nabla_I C_{J \overline{L}}^{\overline{K}} = \nabla_J C_{I \overline{L}}^{\overline{K}}$; complex conjugating gives $\nabla_{\overline I} \overline{C}_{\overline{J} L}^{K} = \nabla_{\overline{J}} \overline{C}_{\overline{I} L}^{K}$, which is the form we are using, with $\Phi$ standing in for the $L$ index.}
\begin{equation}
	(\nabla_{\bar c} \mu_\Phi)_{\bar a}^b = (\nabla_{\bar a} \mu_\Phi)_{\bar c}^b.
\end{equation}
Thus $\mu_\Phi \in \Omega^{0,1}(\cM_{\twod}, T \cM_{\twod})$ is antiholomorphic and 
$\bar\partial$-closed.

The fact that $\mu_\Phi$ is $\bar\partial$-closed means it defines
a class 
\begin{equation}
[\mu_\Phi] \in H^1(\cM_{\twod}, T\cM_{\twod}) = \Def(\cM_{\twod}) 
\end{equation}
(see \autoref{app:beltrami} for a quick review). 	
We claim that
the class $[\mu_\Phi]$ represents
the infinitesimal deformation of $\cM_{\twod}$ induced by perturbing the bulk 
theory with $Q^4 \Phi$.
Indeed, this deformation can be computed by using the Ehresmann
connection in $\cM_{\twodfourd}$ to identify the fibers
with a fixed space $\cM_{\twod}$ to first order in $\eps$, 
and then looking at the $\eps$ dependence of vectors in
$T^{0,1}_\eps \cM_{\twod}$; this is what we have computed in
\eqref{eq:main-derivative}.

The fact that $\mu_\Phi$ is covariantly antiholomorphic constrains the
representative in the class $[\mu_\Phi]$: indeed it
implies that $\mu_\Phi$ can be written in the form
\begin{equation}
	\mu_\Phi = g^{-1} \overline{P_\Phi}
\end{equation}
where $g$ is the the Zamolodchikov metric on $\cM_\twod$ and $P_\Phi$ is a 
holomorphic quadratic differential on $\cM_{\twod}$ (or perhaps meromorphic if $\cM_{\twod}$ has singularities, e.g. the punctures
in class $S$ theories).

\section{Example: \texorpdfstring{$U(1)$}{U(1)} gauge theory with solenoid defect} \label{sec:ex}

In this section we check our results in a very simple example: the free $\N=2$ $U(1)$ gauge theory with a solenoid defect. In \autoref{sec:5.1} we give definitions and fix normalizations. This is followed by a computation of the deformation of the defect moduli space in \autoref{sec:5.2}.

\subsection{Setup} \label{sec:5.1}

We consider the pure $\cN=2$ theory with gauge group $U(1)$, described by a free vector multiplet $(\phi, A_\mu, \lambda_i{}^\alpha, D^{ij})$. Here, $\phi$ is a complex scalar, $A_\mu$ is a $U(1)$ gauge field, $\lambda_i{}^\alpha$ are fermions in the doublets of $\mathfrak{su}(2)_L$ and $\mathfrak{su}(2)_r$, and $D^{ij}$ is an $\mathfrak{su}(2)_r$ triplet of auxiliary fields, 
\bea 
D^{ij} = \begin{pmatrix}
\sqrt{2} F & \I D \\
\I D & \sqrt{2} \overline{F} \\
\end{pmatrix}.
\eea
The action is: 
\begin{equation}
S_{\fourd} = \frac{1}{g^2} \int \de^4 x \left( \frac{1}{4} F_{\mu \nu} F^{\mu \nu} - \phi \Box \overline{\phi} - |F|^2 - \frac{1}{2} D^2 + \I \lambda_i{}^\alpha \sigma^\rho_{\alpha \dot{\beta}} \partial_\rho \overline{\lambda}^{i\dot{\beta}} \right) + \frac{\I \vartheta}{32\pi^2} \int \de^4 x \widetilde{F}^{\mu \nu}F_{\mu \nu}.
\end{equation}

The supersymmetry transformations of the fields are:
\bea \label{eq:rules1}
Q^i{}_\alpha \ \phi &=& \sqrt{2} \lambda^i{}_\alpha \quad \quad \quad \quad \quad \quad \quad \ \ \ \quad ; \quad  \overline{Q}_{j\dot{\beta}} \ \overline{\phi} = \sqrt{2} \overline{\lambda}_{j\dot{\beta}} 
\\ \label{eq:rules2}
Q^i{}_\alpha \ A_{\mu} &=& + i \sigma_{\mu \alpha \dot{\gamma}} \overline{\lambda}^{i\dot{\gamma}} 
\quad \quad \quad \quad \quad \quad \quad   ; \quad 
\overline{Q}_{j\dot{\beta} }\   A_\mu=  i \lambda_j{}^\gamma \sigma_{\mu \gamma \dot{\beta}}   
\\ \label{eq:rules3}
Q^{i}{}_\alpha \overline{\lambda}^{k\dot{\gamma}} &=& - i \sqrt{2} \epsilon^{\dot{\alpha} \dot{\gamma}} \epsilon^{ik} \sigma^\mu{}_{\alpha \dot{\alpha}} \partial_\mu \overline{\phi} 
\quad \quad \quad  ; \quad 
\overline{Q}_{j\dot{\beta}} \lambda_k{}^\gamma = i \sqrt{2} \epsilon_{jk} \epsilon^{\gamma \alpha} \sigma^{\mu}{}_{\alpha \dot{\beta}} \partial_\mu \phi \\ \label{eq:rules4}
Q^i{}_\alpha \ \lambda_j{}^\gamma &=& \delta_\alpha{}^\gamma D_j{}^i - \delta^i{}_j (\epsilon \sigma^{\rho \sigma} \epsilon)^{\gamma}{}_{\alpha} F_{\rho \sigma} \quad
; \quad 
\overline{Q}_{j\dot{\beta}} \ \overline{\lambda}^{i\dot{\gamma}} = - \delta^{\dot{\gamma}}{}_{\dot{\beta}} D_j{}^i + (\epsilon \overline{\sigma}^{\rho \sigma} \epsilon)_{\dot{\beta}}{}^{\dot{\gamma}} \delta_j{}^i F_{\rho \sigma} 
\\ \label{eq:rules5}
Q^{i}{}_\alpha D^{kl} &=& 2 i \epsilon_{\alpha \beta}  \epsilon^{i(l} \overline{\sigma}^{\mu \dot{\beta} \beta} \partial_\mu \overline{\lambda}^{k)}{}_{\dot{\beta}} \quad \quad \quad \ \  ; \quad 
\overline{Q}_{j\dot{\beta}} D^{kl} = 2 i \epsilon_{\dot{\beta} \dot{\rho}} \delta^{(k}{}_j \overline{\sigma}^{\mu \dot{\rho} \beta} \partial_\mu \lambda^{l)}{}_\beta 
\eea
We consider the chiral operator $\Phi = \phi^2$, and its marginal descendant given by:
\bea \label{eq:marg}
Q^4 \phi^2 (x) = \frac{1}{8\pi \I}  \left( - \phi \Box  \overline{\phi} - \I \lambda_i{}^\alpha \sigma^\rho_{\alpha \dot{\beta}} \partial_\rho \overline{\lambda}^{i\dot{\beta}} + \frac{1}{4} F_{\mu \nu} F^{\mu \nu} - \frac{1}{4} \widetilde{F}^{\mu \nu} F_{\mu \nu} - \frac{1}{4} D_{ij} D^{ij}  \right).
\eea
Here $ \widetilde{F}^{\mu \nu} = \frac{1}{2} \epsilon^{\mu \nu \rho \sigma} F_{\rho \sigma}$, and in what follows, we write $F^{\pm} = F \pm \widetilde{F}$. The parameter associated to the marginal operator is the complexified gauge coupling $\tau = \frac{4\pi \I}{g^2} + \frac{\vartheta}{2\pi}$.

We insert a $\half$-BPS defect on $\mathbb{R}_{x^1=x^2=0}$ defined by the action \cite{Gukov:2006jk} 
\bea \label{eq:defectaction}
S_{\twod} &=&- \int\limits_{\mathbb{R}^{2}}  \de^2 x  \left(  \frac{4\pi  \alpha}{g^2} \left(F_{12} +D\right) + \left( \eta+ \frac{\vartheta}{2\pi} \alpha\right) \I F_{34}  \right).
\eea
This defect behaves like a solenoid: its only effect is to create an Aharonov-Bohm phase. 
The first term gives the boundary condition $A = \alpha \, \de \theta$ to the gauge field around $\mathbb{R}^2_{x^1=x^2=0}$. The gauge invariant information is captured by the holonomy $\E^{2\pi \I \alpha}$, and hence $\alpha \sim \alpha + 1$. The second term is the 2d theta term, and hence $\eta \sim \eta + 1$.

The bulk chiral operator $\phi$ and anti-chiral operator $\overline{\phi}$ are also chiral and anti-chiral, respectively, with respect to the defect supersymmetry algebra. The second descendants $Q^2 \phi = \frac{1}{2\sqrt{2}} Q^1{}_1 Q^2{}_2 \phi$ and $\bQ^2 \overline{\phi} = \frac{1}{2\sqrt{2}} \bQ_{2\dot{2}} \bQ_{1\dot{1}}\overline{\phi}$ are marginal operators on the
defect, explicitly given by
\begin{align} 
Q^2 \phi &= \frac{\I}{2}  (D+F_{12} - F_{34}), \\ 
\bQ^2 \overline{\phi} &= - \frac{\I}{2} (D+F_{12} + F_{34}).
\end{align}
The coupling constant associated to $Q^2 \phi$ is the FI parameter
\begin{equation}
 z = \eta + \tau \alpha. 
\end{equation}
This parameter lies in the complex torus $\cM_{\twod} = \mathbb{C}/\Lambda_\tau$, where $\Lambda_\tau = \Z \oplus \tau \Z$. A deformation of the marginal coupling $\tau$ thus induces a deformation of the complex structure on $\cM_{\twod}$.

\subsection{Complex deformation} \label{sec:5.2}

We consider the bulk chiral operator 
$\Phi = \phi^2$ and defect chiral operator $\Sigma = \phi$.
Their OPE is
\begin{equation}
	\Phi(x) \bSigma(0) = - \frac{g^2}{2 \pi^2 \abs{x}^2} \Sigma(0) + \cdots
\end{equation}
Comparing this with \eqref{eq:mu-phi} we have
\begin{equation}
	\mu_\Phi(\bSigma) = \frac{g^2}{8 \pi \I} \Sigma = \frac{1}{\tau - \bar\tau} \Sigma.
\end{equation}
Identifying $\bSigma \to \partial_{\bar z}$ and $\Sigma \to \partial_z$,
this gives $\mu_\Phi \in \Omega^{0,1}(\cM_{\twod}, T \cM_{\twod})$
as
\begin{equation}
	(\mu_\Phi)_{\bar z}^z = \frac{1}{\tau - \bar\tau}.
\end{equation}
This tensor indeed represents the deformation 
of $\cM_{\twod}$ corresponding to $Q^4 \Phi$,
i.e. to $\partial_\tau$.
Indeed, changing $\tau$ to $\tau + \delta \tau$ 
can be realized explicitly
by changing the complex coordinate on $\cM_\twod$ from 
$z$ to $z' = z + \alpha \delta \tau$, and under this coordinate
change we have
\bea 
\frac{\partial}{\partial \overline{z'}} = \frac{\partial}{\partial \overline{z}} +\frac{\delta \tau}{\tau - \overline{\tau}} \frac{\partial}{\partial z}.
\eea
Comparing this with \eqref{eq:Beltramid} we see that
the deformation can be represented by the tensor
$\mu_{\bar z}^z = \frac{1}{\tau - \bar\tau}$,
which indeed matches the $\mu_\Phi$ we computed above.
This verifies that our prescription for the deformation of
$\cM_\twod$ induced by a bulk deformation 
works in this case.

\appendix

\section{Four-dimensional superconformal algebra} \label{app:4d}
The four-dimensional conformal algebra is generated by translation ($P_\mu$), special conformal transformation ($K_\mu$), rotations ($J_{\mu \nu }$) and dilatation ($D$).  When extended to have $\mathcal{N}=2$ supersymmetry, the algebra has eight Poincar\'e supercharges ($Q^{i}{}_\alpha$, $\bQ_{j\dot{\beta}}$) and eight special conformal supercharges ($S_i{}^\alpha$, $\overline{S}^{j\dot{\beta}}$). The latin indices denote the components of $\mathfrak{su}(2)_r$ doublet, and undotted (dotted) Greek indices denote the components of $\mathfrak{su}(2)_L$ ($\mathfrak{su}(2)_R$) doublets. We raise and lower indices of each $\mathfrak{su}(2)$ doublet through the two-dimensional Levi-Civita tensor $\epsilon_{i j}$, where we use the convention $\epsilon_{21} = \epsilon^{12} = 1$.  The generators of $\mathfrak{su}(2)_L \oplus \mathfrak{su}(2)_R$ are related to $J_{\mu \nu}$ through the following relations.\footnote{Our conventions compare with the conventions of \cite{Dolan:2002zh} as follows: for $x \in \{\ovS^{i\dot{\alpha}} , K_\mu , \overline{J}^{\dot{\alpha}}{}_{\dot{\beta}} \}$, $x^{\text{here}} = - x^{\text{there}}$. Also, $D^{\text{here}} = -i D^{\text{there}}$.} 
\bea 
J_{\alpha}{}^\beta &=& - \frac{i}{4} (\sigma^\mu \overline{\sigma}^\nu)_{\alpha}{}^\beta J_{\mu \nu} ; \quad
\overline{J}^{\dot{\alpha}}{}_{\dot{\beta}} =  \frac{i}{4} (\overline{\sigma}^\mu \sigma^\nu)^{\dot{\alpha}}{}_{\dot{\beta}} J_{\mu \nu}
\eea
The nonzero anti--commutators of odd generators are as follows. 
\bea 
\{Q^i{}_\alpha , \bQ_{j\dot{\beta}} \} &=& 2 \delta^i{}_j \ \sigma^\mu_{\alpha \dot{\beta}} \ P_\mu  \\ 
\{S_i{}^\alpha , \ovS^{j\dot{\beta}} \} &=& 2 \delta_i{}^j \ \overline{\sigma}^{\mu \dot{\beta} \alpha} \ K_\mu \\
\{Q^i{}_\alpha , S_j{}^\beta \} &=& 4 \left(\delta^i{}_j J_\alpha{}^\beta - \delta_\alpha{}^\beta R^i{}_j + \frac{1}{2} \delta^i{}_j \delta_\alpha{}^\beta D \right) \\
\{\bQ_{j\dot{\beta}}, \ovS^{i\dot{\alpha}} \} &=& 4 \left( \delta^i{}_j \overline{J}^{\dot{\alpha}}{}_{\dot{\beta}} + \delta^{\dot{\alpha}}{}_{\dot{\beta}} R^i{}_j + \frac{1}{2} \delta^i{}_j \delta^{\dot{\alpha}}{}_{\dot{\beta}} D  \right)
\eea

Here and throughout the paper, $\sigma^\mu = (\vec{\sigma}, i)$, $\overline{\sigma}^\mu = (-\vec{\sigma}, i)$. The components of $\vec{\sigma}$ are Pauli matrices. 
\bea 
[J_{\alpha}{}^\beta, Q^{i}{}_\gamma] &=& \delta_{\gamma}{}^\beta Q^i{}_\alpha - \frac{1}{2} \delta_\alpha{}^\beta Q^i{}_\gamma \\
{ [ \overline{J}^{\dot{\alpha}}{}_{\dot{\beta}} , \bQ_{j\dot{\gamma}} ]}  &=& \delta^{\dot{\alpha}}{}_{\dot{\gamma}} \bQ_{j\dot{\beta}} - \frac{1}{2} \delta^{\dot{\alpha}}{}_{\dot{\beta}} \bQ_{j\dot{\gamma}}  \\
{ [J_{\alpha}{}^\beta, S_{i}{}^\gamma] } &=& 
- \delta_{\alpha}{}^\gamma S_i{}^\beta + \frac{1}{2} \delta_\alpha{}^\beta S_i{}^\gamma \\
{[\overline{J}^{\dot{\alpha}}{}_{\dot{\beta}} , \ovS^{i\dot{\gamma}}] } &=& - \delta^{\dot{\gamma}}{}_{\dot{\beta}} \ovS^{i\dot{\alpha}} + \frac{1}{2} \delta^{\dot{\alpha}}{}_{\dot{\beta}} \ovS^{i\dot{\gamma}}
\eea
The nonzero commutators of supercharges with translations and special conformal transformations are as follows.
\bea
{[K_\mu , Q^{i}{}_\alpha ]} &=& - (\sigma_\mu)_{\alpha\dot{\alpha}} \overline{S}^{i\dot{\alpha}} \quad \quad,  \quad \quad 
{[K_\mu , \bQ_{j\dot{\beta}}]} = -(\sigma_\mu)_{\alpha \dot{\beta}}  \\
{ [P_\mu , \overline{S}^{i\dot{\alpha}} ] } &=&  (\overline{\sigma}_\mu)^{\dot{\alpha} \beta}  Q^i{}_{\beta} \quad \quad \ \ ,  \quad \quad
{ [P_\mu , S_i{}^\alpha ] } =  (\overline{\sigma}_\mu)^{\dot{\alpha} \alpha}  \bQ_{i\dot{\alpha}}
\eea
The $\mathfrak{su}(2)_r$ algebra has ladder operators $R_{\pm}$ and Cartan generator $R_3$. The supercharges $Q^1{}_\alpha$, $\bQ_{2\dot{\alpha}}$, $S_2{}^\alpha$ and $\ovS^{1\dot{\alpha}}$ have $R_3$-charge $+\frac{1}{2}$; while $Q^2{}_\alpha$, $\bQ_{1\dot{\alpha}}$, $S_1{}^\alpha$ and $\ovS^{2\dot{\alpha}}$ have $R_3$-charge $-\frac{1}{2}$. Lastly, the scaling dimensions and $\mathfrak{u}(1)_r$ charges of all superchargess are as follows. 
\bea 
{[ D, Q^{i}{}_\alpha ]} &=&[r, Q^{i}{}_\alpha] = \frac{1}{2} Q^{i}{}_\alpha \\
- {[D, \bQ_{j\dot{\beta}} ]} &=&  {[r,\bQ_{j\dot{\beta}} ]} = - \frac{1}{2} \bQ_{j\dot{\beta}} \\
{ [D, S_i{}^\alpha ] } &=& { [r, S_i{}^\alpha ] }  =  - \frac{1}{2} S_i{}^\alpha \\
- { [ D, \ovS^{i\dot{\alpha}} ] } &=& { [ r, \ovS^{i\dot{\alpha}}] } =  \frac{1}{2} \ovS^{i\dot{\alpha}} 
\eea

\section{Two-dimensional superconformal algebra} \label{app:2d}
Now we discuss the superconformal algebra in two dimensions with $\mathcal{N}=(2,2)$ supersymmetry. We take the surface defect to be present at $x^1=x^2=0$. The conformal algebra in two dimensions in terms of holomorphic operators ($L_n$) and anti-holomorphic operators ($\overline{L}_n$) for $n=-1$, $0$, $ +1$.
\bea 
i P_4 &=& L_{-1} + \overline{L}_{-1}, \quad  \quad 
P_3 = L_{-1} -\overline{L}_{-1} \\
i K_4 &=& L_{1} + \overline{L}_{1}, \quad \quad \quad \
K_3 = \overline{L}_1 - L_{1}\\
D &=&  L_0 + \overline{L}_0 , \quad \quad \quad \
J_{34} =  L_0 - \overline{L}_0
\eea
In an $\mathcal{N}=(2,2)$ superconformal algebra, there are eight fermionic generators which we notate as $G^{\pm}_{1/2}$, $G^{\pm}_{-1/2}$, $\overline{G}^{\pm}_{1/2}$ and $\overline{G}^{\pm}_{-1/2}$. The nonzero (anti-) commutation relations of the operators are as follows.
\bea 
[L_n , G^{\pm}_r] &=& \left( \frac{n}{2} - r \right) G^{\pm}_{n+r} \\
{[\overline{L}_n, \bG^{\pm}_r]} &=&  \left( \frac{n}{2} - r \right) \overline{G}^{\pm}_{n+r} \\
\{G^+_{r}, G^-_{s} \} &=& L_{r+s} + \frac{r-s}{2} J_{r+s} \\
\{\bG^+_{r}, \bG^-_{s} \} &=& \overlineL_{r+s} + \frac{r-s}{2} \bJ_{r+s}
\eea
The algebra has $\mathfrak{u}(1)_L \oplus \mathfrak{u}(1)_R $ R-symmetry with generators $J_0$, $\bJ_0$.
\bea 
[J_0 , G^a_{\pm 1/2}] &=& a G^a_{\pm 1/2}, \quad \quad [\bJ_0 , \bG^a_{\pm 1/2}] = a \bG^a_{\pm 1/2}
\eea
Comparison with four-dimensional $\mathcal{N}=2$ superconformal algebra gives the following identification of operators.
\bea \label{eq:B.9}
G^+_{-1/2} &=& \frac{Q^1{}_1}{2} ,  \quad G^-_{-1/2} = \frac{\bQ_{\dot{1}1}}{2} , \quad G^-_{+1/2} = \frac{S_1{}^1}{2} , \quad G^+_{+1/2} =  \frac{\overline{S}^{1\dot{1}}}{2} \\ \label{eq:B.10}
\overline{G}^-_{-1/2} &=& \frac{Q^2{}_2}{2} ,  \quad \overline{G}^+_{-1/2} = \frac{\bQ_{\dot{2}2}}{2} , \quad \overline{G}^+_{+1/2} = \frac{S_2{}^2}{2} , \quad \overline{G}^-_{+1/2} = \frac{\overline{S}^{2\dot{2}}}{2} \\
J_0 &=& 2R_3 + J_{12} + r , \quad \quad \quad \ \ \
\overline{J}_0 =  2R_3 + J_{12} - r 
\eea

We choose the following conventions for the vector and axial R-charges: $J_V = J_0 - \bJ_0$ and $J_A =  J_0 + \overline{J}_0$. With this convention, the bulk chiral primary is also a chiral primary with respect to the defect superconformal algebra.

\section{Deformations of complex manifolds} \label{app:beltrami}

Here we review some basic facts about deformations of complex manifolds; see e.g.
 \cite{huybrechts_2005} for a more complete treatment.

Let $M$ be a manifold of real dimension $2n$. 
A complex structure on $M$ is a section $I \in \End(TM)$ such that $I^2 = -1$ and the Nijenhuis tensor $N_I$ vanishes. When the latter condition is not necessarily satisfied, $I$ is an almost complex structure on $M$. 

The data of an almost complex structure $I$ is uniquely determined by the decomposition 
\begin{equation}
T_\mathbb{C} M = T^{1,0} M \oplus T^{0,1} M
\end{equation}
such that $I\vert_{T^{1,0} M} = \I$ and $I\vert_{T^{0,1} M} = -\I$. 
Now suppose we have a family of almost complex structures 
$I_\eps$ on $M$, parameterized by a variable $\epsilon$. Then we have a family of decompositions with respect to $I_\eps$,
\bea 
T_\mathbb{C} M = T^{1,0}_\epsilon M \oplus T^{0,1}_{\epsilon} M.
\eea
For sufficiently small $\epsilon$, the deformation $I_\epsilon$ may be encoded by a map
\bea \label{eq:map}
J_\epsilon : T^{0,1}_0 M  \to T^{1,0}_0 M
\eea
such that if $v \in T^{0,1}_0 M$ then $v + \epsilon J_\epsilon v \in T^{0,1}_\epsilon M$. In this paper, we will only be interested in studying deformations of complex structure to first order in $\eps$. To this order, the requirement that the Nijenhuis tensor of $I_\epsilon$ vanishes is equivalent to the requirement that $J_\epsilon \in \Omega^{0,1}_0 (M, TM)$ is $\overline{\partial}$-closed.

We say that two complex structures are isomorphic if there exists a self-diffeomorphism of $M$ that maps one to the other. Given a one-parameter family of diffeomorphisms $F_\epsilon$, we obtain a global vector field on $M$, $\frac{\de F_\epsilon}{\de \epsilon} \in \Omega^0 (TM)$. Conversely, given a vector field on $M$, we get an infinitesimal diffeomorphism of $M$. Pushing the complex structure $I$ forward through $F_\epsilon$ gives a new complex structure $I_\epsilon = \de F_\epsilon \circ I (\de F_\epsilon)^{-1}$, isomorphic to the original one, with the first-order deformation
\bea 
J_\eps = \overline{\partial} \left( \left(\frac{\de F_\epsilon}{\de \epsilon} \right)^{1,0} \right).
\eea
Such a $J_\eps$ is thus regarded as a trivial first-order deformation. 
Thus altogether
the space $\Def(X)$ of first-order deformations of a complex manifold 
$X$ is isomorphic to the Dolbeault cohomology $H^{0,1} (X, TX)$. 

In local coordinates on $X$, a representative $\mu$ may be written as 
\bea 
\mu = \sum\limits_{a,b=1}^n \mu_{\bar a}^b \de \overline{z}^a \frac{\partial}{\partial z^b}.
\eea
Through the map (\ref{eq:map}) the antiholomorphic tangent vectors in the deformed complex structure $I_\epsilon$ are given as
\bea  \label{eq:Beltramid}
\left(\frac{\partial}{\partial \overline{z}^a} \right)_\epsilon = \frac{\partial}{\partial \overline{z}^a} + \epsilon \mu_{\bar a}^b \frac{\partial}{\partial z^b}.
\eea

\section{The covariant derivative of the supercurrent} \label{sec:deriv-J}

\subsection{Computation} \label{sec:deriv-J-computation}

In this section we give the computation leading to \eqref{eq:supercurrent-variation}.
To simplify notation here we suppress some indices, writing $\bJ^\mu$ for the supercurrent 
$\bJ^\mu_{I \dot\alpha}$.

The idea of the computation is a bit indirect.
We first consider the extension of $\bJ^\mu$ to a section $\bJ^\mu_\cc$
of the bundle of operators over the family of deformed theories, to first order in $\eps$,
obeying $\nabla_\eps \bJ^\mu_\cc = 0$.
The key point is that $\bJ^\mu_\cc$ is not an allowed supercurrent, 
because it is not conserved at $\eps \neq 0$.
We calculate the amount by which conservation of $\bJ^\mu_\cc$ is violated, by
writing out the inner product 
\begin{equation}
	\partial_\mu \IP{\Psi(\infty) \bJ^\mu_\cc(x)}
\end{equation}
between $\partial_\mu \bJ^\mu_\cc$ and a general 
operator $\Psi$ that is also extended in a covariantly constant manner.
To first order in $\epsilon$
we have
\begin{equation}
\partial_\mu \IP{\Psi(\infty) \bJ^\mu_\cc(x)} = - \eps \partial_\mu \int \de^4 y \IP{\Psi(\infty) \Phi^{(4)}(y) \bJ^\mu(x)}
\end{equation}
Formally this would seem to be zero because we can change variables 
to $y' = y - x$ in the integral, so that the integral is
$x$-independent. But we will argue now that it is not actually zero
once we take care of regularizations.

For this we need the
OPE between $\Phi^{(4)}$ and $\bJ^\mu$:
\begin{equation}
\bJ^\mu(x) \Phi^{(4)}(0) = \frac{1}{2 \pi^2} \frac{x^\mu}{\abs{x}^4} \bQ \Phi^{(4)}(0) + \cdots  
\end{equation}
where $\cdots$ denotes operators of dimension different from $\frac92$, and again we suppress indices,
writing $\bQ$ for $\bQ_{I \dot\alpha}$. 
Inserting this OPE in our integral gives
\begin{equation}
\partial_\mu \IP{\Psi(\infty) \bJ_\cc^\mu(x)} = \frac{1}{2\pi^2} \eps \partial_\mu \int \de^4 y \frac{(y-x)^\mu}{\abs{y-x}^4} \IP{\Psi(\infty) \bQ \Phi^{(4)}(x)} + \cdots
\end{equation}
The domain $R_x$ of integration here is a big ball $\abs{y} < M$, 
excluding a small ball around
$y = x$. Changing variable to $y' = y-x$, the domain $R'_x$ becomes a ball
$\abs{y'+x} < M$ with a small ball around $y' = 0$ deleted, and
\begin{equation}
\partial_\mu \IP{\Psi(\infty) \bJ^\mu_\cc(x)} = \frac{1}{2\pi^2} \eps \partial_\mu \left[ \IP{\Psi(\infty) \bQ \Phi^{(4)}(x)} \int_{R'_x} \de^4 y' \frac{y'^\mu}{\abs{y'}^4} \right] + \cdots
\end{equation}
If we specialize to the case where $\Psi$ is an operator of dimension $\frac92$, 
then the extra contributions $\cdots$ vanish, and also we can replace
$x$ by $0$ in the vev, getting
\begin{equation}
\partial_\mu \IP{\Psi(\infty) \bJ^\mu_\cc(x)} = \frac{1}{2\pi^2} \eps \IP{\Psi \vert \bQ \Phi^{(4)}} \, \partial_\mu  \left[ \int_{R'_x} \de^4 y' \frac{y'^\mu}{\abs{y'}^4} \right]
\end{equation}
Differentiating the integral gives $-2 \pi^2$ (see next section) and we get finally
\begin{equation} \label{eq:non-conservation}
\partial_\mu \bJ^\mu_\cc = - \eps \bQ \Phi^{(4)}.
\end{equation}

How do we interpret \eqref{eq:non-conservation}? It means that
$\bJ^\mu_\cc$ is not conserved, but it also shows us how to fix the
problem: if we find an operator
$\cO^\mu$ obeying
\begin{equation} \label{eq:O-condition}
 \partial_\mu \cO^\mu = \bQ \Phi^{(4)},
\end{equation}
then we could consider a new section 
\begin{equation}
\bJ^\mu_\cons = \bJ^\mu_\cc + \eps \cO^\mu
\end{equation}
and then $\bJ^\mu_\cons$ \ti{is} a conserved supercurrent, obeying\footnote{The derivative $\nabla_\epsilon$ is computed at $\epsilon=0$. Therefore, we do not write down $\epsilon \nabla_\epsilon \cO^\mu$ in equation (\ref{eq:covsc}). }
\begin{equation} \label{eq:covsc}
\nabla_\eps \bJ^\mu_\cons = \cO^\mu.
\end{equation}
Moreover, the condition \eqref{eq:O-condition} has a canonical solution: we can 
take $\cO = \Phi^{(3)}$.
Any other solution must differ from $\Phi^{(3)}$ by a conserved supercurrent,
which can only be of the form $c \bJ^\mu$ for some scalar $c$ (since we 
assume the theory has exactly $\cN=2$ supersymmetry, not more).
Thus finally
\begin{equation}
\nabla_\eps \bJ^\mu_\cons = (\Phi^{(3)})^\mu + c \bJ^\mu
\end{equation}
which is the desired \eqref{eq:supercurrent-variation}.

\subsection{An integral}

In this section we relabel $y' \to y$ and $R' \to R$,
and we work in $d$ dimensions.
Fix parameters $M, \eps$ and define the region 
\begin{equation}
	R_x = \{ y: \abs{y+x} < M, \abs{y} > \eps \}.
\end{equation}
Then let
\begin{equation}
I^\mu(x) = \int_{R_x} \de^d y \frac{y^\mu}{\abs{y}^d}.
\end{equation}
For a fixed $\mu$,
the derivative $\partial_\mu I^\mu$ can be computed as follows.
We shift $x$ by (an infinitesimal) $t$ units in the positive $\mu$ direction. 
Consider the side $y^\mu > 0$ of $R_x$. On this side, the size of $R_x$ is decreased
by translating the frontier $y$ by $-t$ in the $\mu$ direction.
On the other side we are similarly increasing the size of $R_x$.
On each side we can approximate the integrand by its value on the sphere.

The change in $I^\mu$ can thus be represented as an integral over the coordinates other than $\mu$,
which span a ball $B^{d-1}$, with radial coordinate $r$ and the usual area element $\de A_{d-1}$:
\begin{align}
\delta I^\mu &= 2 \int_{B^{d-1}_M} \frac{\sqrt{M - r^2}}{M^d}  (-t) \, \de A_{d-1}  \\
&= (-t M^{-d}) \int_0^M \de r \int_{S^{d-2}_r} 2\sqrt{M - r^2} \, \de A_{d-1} \\
&= (-t M^{-d}) \Vol(B^d_M) \\
&= -t \Vol(B^d)
\end{align}
Thus we have with $\mu$ fixed
\begin{equation}
  \partial_\mu I^\mu = - \Vol(B^d)
\end{equation}
and summing over the index $\mu$,
\begin{equation}
  \partial_\mu I^\mu = - d \Vol(B^d) = - \Vol(S^{d-1}).
\end{equation}

\subsection{With a surface defect} \label{sec:deriv-J-with-defect}

In the presence of a \half-BPS surface defect $\SD(P)$,
in addition to the bulk supercurrent $\bJ^\mu$ there is also the 
defect supercurrent $\bJ_\defect^\mu$, defined 
on $P$, with the property that
(loosely speaking) 
the combined $\bJ + \delta(P) \bJ_\defect$ is conserved.
One sharp way to express this conservation is to say that
 $\partial_\mu \bJ^\mu_\defect$ has to cancel the delta-function
 divergence of
the bulk supercurrent $\bJ^\mu$ in the normal directions, which
we can measure as the flux through a small circle linking
the defect: the pair $(\bJ, \bJ_\defect)$ has to obey
\begin{equation} \label{eq:combined-conservation}
	\partial_\mu \bJ^\mu_\defect(y) = \lim_{r \to 0} r \oint_{S^1_r(y)} \bJ^r(x) \, \de \theta \, .
\end{equation}

Now, in parallel to the discussion of
$\nabla_\eps \bJ^\mu$ in \autoref{sec:deriv-J-computation}, 
we may consider the
covariant derivative $\nabla_\eps \bJ^\mu_\defect$.
A computation parallel to that of 
\autoref{sec:deriv-J-computation} gives no order $\eps$ 
contribution to $\partial_\mu \bJ^\mu_{\defect,\cc}$; 
thus to first order in $\eps$ we have the same equation as at $\eps = 0$,
\begin{equation}
	\partial_\mu \bJ^\mu_{\defect,\cc}(y) = \lim_{r \to 0} r \oint_{S^1_r(y)} \bJ^r_\cc(x) \, \de \theta.
\end{equation}
Replacing $\bJ^r_\cc$ in favor of the bulk conserved supercurrent
$\bJ^r_\cons$ this becomes
\begin{equation}
	\partial_\mu \bJ^\mu_{\defect,\cc}(y) = \lim_{r \to 0} r \oint_{S^1_r(y)} \left( \bJ^r_\cons(x) - \eps (\Phi^{(3)})^r \right) \de \theta.
\end{equation}
Thus, if we define $\bJ^\mu_{\defect,\cons}$ by
\begin{equation}
	\bJ^\mu_{\defect,\cons} = \bJ^\mu_{\defect,\cc} + \eps C^\mu, 
\end{equation}
where $C^\mu$ is a defect operator obeying
\begin{equation} \label{eq:C-condition}
	\partial_\mu C^\mu(y) = r \lim_{r \to 0} \oint_{S^1_r(y)} (\Phi^{(3)})^r \de \theta \, ,
\end{equation}
then the pair $(\bJ^\mu_\cons, \bJ^\mu_{\defect,\cons})$
obeys the desired \eqref{eq:combined-conservation}.

To determine $C^\mu$ 
we now examine the right side of \eqref{eq:C-condition}.
It is nonzero just if there is a divergence of order $1/r$
in the OPE between $(\Phi^{(3)})^r$
and the surface defect $\SD(P)$. 
For notational convenience let us consider just the single
component $(\Phi^{(3)})^r_{2 \dot 2}$.
Using \eqref{eq:phi3-r} 
and the fact that $\SD(P)$ preserves $Q^1{}_1$
and $Q^2{}_2$, this OPE takes the form
\begin{equation}
(\Phi^{(3)})^r_{2 \dot{2}} \SD(P) = \E^{-\I \theta} Q^1{}_1 Q^2{}_2 \left(Q^1{}_2 \Phi(x) \SD(P) \right).
\end{equation}
$\Phi(x)$ has $\Delta = 2$, $r = -2$, 
and thus $Q^1{}_2 \Phi(x)$ has $\Delta = \frac52$, $r = -\frac32$.
An operator with this $R$-charge must have dimension at least
$\frac32$;
the most singular term in the OPE is thus of the form
\begin{equation} \label{eq:most-sing}
	(\Phi^{(3)})^r_{2 \dot{2}} \SD(P) = \frac{1}{r} Q^1{}_1 Q^2{}_2 \left( \xi(y) \SD(P) \right) + \cdots
\end{equation}
where $y$ is the projection of $x$ to $P$.
The defect operator $\xi$ saturates $R$-charge bounds on both sides
(in the 2-d SCFT notation it has $(L_0, j_0) = (\half, -1)$ and $(\overline{L}_0, \overline{j}_0) = (1, 2)$), and thus 
it is a superconformal primary.
Combining \eqref{eq:most-sing} with \eqref{eq:C-condition} we have
\begin{equation}
	\partial_\mu C^\mu = 2 \pi Q^1_1 Q^2_2 \xi.
\end{equation}
However, the fact that $\xi$ is a superconformal primary
implies $Q^1{}_1 Q^2{}_2 \xi$ is orthogonal to all conformal
descendants; thus we conclude that $Q^1{}_1 Q^2{}_2 \xi = 0$.
It follows that we can take $C^\mu = 0$, i.e.
$\nabla_\eps \bJ^\mu_{\defect,\cons} = 0$.

\providecommand{\href}[2]{#2}\begingroup\raggedright\endgroup


\begin{thebibliography}{10}

\bibitem{Alday:2009fs}
L.~F. Alday, D.~Gaiotto, S.~Gukov, Y.~Tachikawa, and H.~Verlinde, ``{Loop and
  surface operators in {N=2} gauge theory and Liouville modular geometry},''
  {\em JHEP} {\bf 01} (2010) 113,
\href{http://www.arXiv.org/abs/0909.0945}{{\tt 0909.0945}}.

\bibitem{Gaiotto:2009fs}
D.~Gaiotto, ``{Surface Operators in N = 2 4d Gauge Theories},'' {\em JHEP} {\bf
  11} (2012) 090,
\href{http://www.arXiv.org/abs/0911.1316}{{\tt 0911.1316}}.

\bibitem{Gaiotto:2009we}
D.~Gaiotto, ``{N=2 dualities},'' {\em JHEP} {\bf 08} (2012) 034,
\href{http://www.arXiv.org/abs/0904.2715}{{\tt 0904.2715}}.

\bibitem{Seiberg:1988pf}
N.~Seiberg, ``{Observations on the Moduli Space of Superconformal Field
  Theories},'' {\em Nucl. Phys.} {\bf B303} (1988)
286--304.

\bibitem{Kutasov:1988xb}
D.~Kutasov, ``{Geometry on the Space of Conformal Field Theories and Contact
  Terms},'' {\em Phys. Lett.} {\bf B220} (1989)
153--158.

\bibitem{Ranganathan:1992nb}
K.~Ranganathan, ``{Nearby CFTs in the operator formalism: The Role of a
  connection},'' {\em Nucl. Phys.} {\bf B408} (1993) 180--206,
\href{http://www.arXiv.org/abs/hep-th/9210090}{{\tt hep-th/9210090}}.

\bibitem{Ranganathan:1993vj}
K.~Ranganathan, H.~Sonoda, and B.~Zwiebach, ``{Connections on the state space
  over conformal field theories},'' {\em Nucl. Phys.} {\bf B414} (1994)
  405--460,
\href{http://www.arXiv.org/abs/hep-th/9304053}{{\tt hep-th/9304053}}.

\bibitem{1812.11199}
V.~V. Fock and A.~Thomas, ``Higher complex structures,''
  \href{http://www.arXiv.org/abs/arXiv:1812.11199}{{\tt arXiv:1812.11199}}.

\bibitem{Papadodimas:2009eu}
K.~Papadodimas, ``{Topological Anti-Topological Fusion in Four-Dimensional
  Superconformal Field Theories},'' {\em JHEP} {\bf 08} (2010) 118,
\href{http://www.arXiv.org/abs/0910.4963}{{\tt 0910.4963}}.

\bibitem{Moore:1993zc}
G.~W. Moore, ``Finite in all directions,''
\href{http://www.arXiv.org/abs/hep-th/9305139}{{\tt hep-th/9305139}}.

\bibitem{Baggio:2014ioa}
M.~Baggio, V.~Niarchos, and K.~Papadodimas, ``{tt$^{*}$ equations, localization
  and exact chiral rings in 4d $ \mathcal{N} $ =2 SCFTs},'' {\em JHEP} {\bf 02}
  (2015) 122, \href{http://www.arXiv.org/abs/1409.4212}{{\tt 1409.4212}}.

\bibitem{Gerchkovitz:2016gxx}
E.~Gerchkovitz, J.~Gomis, N.~Ishtiaque, A.~Karasik, Z.~Komargodski, and S.~S.
  Pufu, ``{Correlation Functions of Coulomb Branch Operators},'' {\em JHEP}
  {\bf 01} (2017) 103, \href{http://www.arXiv.org/abs/1602.05971}{{\tt
  1602.05971}}.

\bibitem{Baggio:2017aww}
M.~Baggio, V.~Niarchos, and K.~Papadodimas, ``{Aspects of Berry phase in
  QFT},'' {\em JHEP} {\bf 04} (2017) 062,
  \href{http://www.arXiv.org/abs/1701.05587}{{\tt 1701.05587}}.

\bibitem{Niarchos:2018mvl}
V.~Niarchos, ``{Geometry of Higgs-branch superconformal primary bundles},''
  {\em Phys. Rev. D} {\bf 98} (2018), no.~6, 065012,
  \href{http://www.arXiv.org/abs/1807.04296}{{\tt 1807.04296}}.

\bibitem{Cecotti:1991me}
S.~Cecotti and C.~Vafa, ``Topological-antitopological fusion,'' {\em Nucl.
  Phys.} {\bf B367} (1991)
359--461.

\bibitem{Gukov:2006jk}
S.~Gukov and E.~Witten, ``{Gauge Theory, Ramification, And The Geometric
  Langlands Program},'' \href{http://www.arXiv.org/abs/hep-th/0612073}{{\tt
  hep-th/0612073}}.

\bibitem{Dolan:2002zh}
F.~A. Dolan and H.~Osborn, ``{On short and semi-short representations for
  four-dimensional superconformal symmetry},'' {\em Annals Phys.} {\bf 307}
  (2003) 41--89,
\href{http://www.arXiv.org/abs/hep-th/0209056}{{\tt hep-th/0209056}}.

\bibitem{huybrechts_2005}
D.~Huybrechts, {\em Complex Geometry - An Introduction}.
\newblock Universitext. Springer-Verlag Berlin Heidelberg, 2005.

\end{thebibliography}
\end{document}